\documentclass[aps, prd, amsmath, floats, floatfix, twocolumn, nofootinbib, superscriptaddress]{revtex4-1}
\usepackage[pdftex]{graphicx}
\usepackage{color}
\usepackage{latexsym}
\usepackage[colorlinks]{hyperref}

\newcommand{\be}{\begin{eqnarray}}
\newcommand{\ee}{\end{eqnarray}}

\begin{document}

\title{XSPEC model for testing the Kerr black hole hypothesis using the continuum-fitting method}

\author{Menglei~Zhou}
\affiliation{Center for Field Theory and Particle Physics and Department of Physics, Fudan University, 200438 Shanghai, China}

\author{Askar~B.~Abdikamalov}
\affiliation{Center for Field Theory and Particle Physics and Department of Physics, Fudan University, 200438 Shanghai, China}

\author{Dimitry~Ayzenberg}
\affiliation{Center for Field Theory and Particle Physics and Department of Physics, Fudan University, 200438 Shanghai, China}

\author{Cosimo~Bambi}
\email[Corresponding author: ]{bambi@fudan.edu.cn}
\affiliation{Center for Field Theory and Particle Physics and Department of Physics, Fudan University, 200438 Shanghai, China}

\author{Honghui~Liu}
\affiliation{Center for Field Theory and Particle Physics and Department of Physics, Fudan University, 200438 Shanghai, China}

\author{Sourabh~Nampalliwar}
\affiliation{Theoretical Astrophysics, Eberhard-Karls Universit\"at T\"ubingen, 72076 T\"ubingen, Germany}

\begin{abstract}
We present {\sc nkbb}, an XSPEC model for the thermal spectrum of thin accretion disks in parametric black hole spacetimes. We employ the Novikov-Thorne model for the description of the accretion disk and the formalism of the transfer function proposed by Cunningham for storing all the relativistic effects of the spacetime metric. The current version of the model assumes the Johannsen metric, but it can be easily extended to any stationary, axisymmetric, and asymptotically-flat spacetime without pathological properties. The model can be used within the XSPEC package to test the Kerr nature of astrophysical black holes with the continuum-fitting method.
\end{abstract}

\maketitle

%%%%%%%%%%%%%%%%%%%%%%%%%%%%%%%

\section{Introduction}

Einstein's theory of general relativity was proposed over a century ago and has successfully passed a large number of observational tests~\cite{will}. However, there are also a number of arguments suggesting new physics, ranging from the problem of spacetime singularities in physically relevant solutions to the black hole information paradox, from the observed accelerated expansion rate of the Universe to the problem of finding a quantum theory of gravity. While the theory is relatively well tested in weak gravitational fields (even if definitively not with the precisions possible in atomic and particle physics), the strong field regime is still largely unexplored\footnote{The recent detection of gravitational waves from the coalescence of black hole binaries can explore the dynamical strong field regime, but the available data do not permit to put significant constraints on possible deviations from general relativity~\cite{2-gw0,2-gw1,2-gw2}. More stringent tests will be presumably possible with space-based gravitational wave detectors~\cite{2-gw3}.}. Astrophysical black hole systems are an ideal laboratory for testing Einstein's gravity in the strong field regime.

The only stationary, axisymmetric, and asymptotically flat, vacuum black hole solution of the 4-dimensional Einstein's equations, regular on and outside the event horizon, is the Kerr metric~\cite{k1,k2,k3}. The spacetime around astrophysical black holes formed from the complete gravitational collapse of some progenitor body should be well approximated by the Kerr solution. Initial deviations from the Kerr metric are indeed expected to be quickly radiated away by the emission of gravitational waves~\cite{k4}. Gravitational fields produced by accretion disks or nearby stars have a negligible impact on the spacetime metric near the black hole event horizon~\cite{k5,k6}. Non-vanishing electric charges can be safely ignored~\cite{k7}. In the end, we should expect that the metric of the spacetime around an astrophysical black hole is very close to the ideal Kerr solution of general relativity, and macroscopic deviations may only be possible in the presence of new physics.

The Kerr black hole hypothesis can be tested by studying the properties of the electromagnetic radiation emitted by the gas of the inner part of the accretion disk or by bodies orbiting the compact object. There are a number of electromagnetic techniques that have been proposed to probe the metric around black holes~\cite{r1,r2,ccc1,ccc2}. As of now, the two leading methods are the analysis of the thermal spectrum of the disk (continuum-fitting method)~\cite{cfm1,cfm2,cfm3} and the study of the reflection spectrum of the disk (X-ray reflection spectroscopy)~\cite{i1,i2}. Both techniques were originally proposed to measure black hole spins under the assumption that the spacetime metric is described by the Kerr solution and more recently they have been suggested as a tool to test the Kerr metric~\cite{nk1,nk2,nk3,nk4,nk5,nk6,nk7,nk8,nk9,nk10,nk11}.

If we want to construct non-Kerr models for the thermal and reflection components of accretion disks and then analyze X-ray data to constrain the spacetime metric around black holes, it is convenient that these models can be used with standard X-ray data analysis packages like XSPEC~\cite{arnaud}. In Refs.~\cite{noi1,noi2}, we presented the XSPEC reflection model {\sc relxill\_nk}, which is an extension of the {\sc relxill} family~\cite{rx1,rx2} to non-Kerr spacetimes. The constraints obtained with {\sc relxill\_nk} from the analysis of specific sources have been reported in~\cite{c1,c2,c3,c4,c5,c6,c7,c8,c9}. In this paper, we present {\sc nkbb} (for Non-Kerr BlackBody), which is an XSPEC model for the thermal spectrum of thin accretion disks to test the Kerr black hole hypothesis using the continuum-fitting method. While there are already some models for the thermal spectrum in phenomenological metrics in which possible deviations from the Kerr solution are described by a number of ``deformation parameters'' (parametric black hole spacetimes), the constraints reported in the literature with these models are qualitative and based on simulations only, see e.g.~\cite{nk8}. {\sc nkbb} is instead an XSPEC compatible model that can be immediately used to test the Kerr metric with real X-ray data.

The content of the paper is as follows. In Section~\ref{s-nt}, we briefly review the Novikov-Thorne model for the description of infinitesimally thin accretion disks in generic stationary, axisymmetric, and asymptotically flat spacetimes. In Section~\ref{s-trans}, we describe the formalism of the transfer function to store all the relevant details about the spacetime metric into a FITS file ready to be used by {\sc nkbb}. In Section~\ref{s-xspec}, we present the current version of {\sc nkbb} and, in Section~\ref{s-comp}, we compare {\sc nkbb} with {\sc kerrbb}~\cite{kerrbb} to verify the accuracy of our model. Section~\ref{s-sim} shows the results of the potential capabilities of {\sc nkbb} to test the spacetime metric using the continuum-fitting method with some simulations. Section~\ref{s-dis} is for the conclusions. Throughout the paper, we employ units in which $G_{\rm N} = c = 1$ and the convention of a metric with signature $(-+++)$.

%%%%%%%%%%%%%%%%%%%%%%%%%%%%%%%

\section{Novikov-Thorne model \label{s-nt}}

The standard framework for the description of geometrically thin and optically thick accretion disks in stationary, axisymmetric, and asymptotically flat spacetimes is the Novikov-Thorne model~\citep{nt1,nt2}. The disk is assumed infinitesimally thin on the plane perpendicular to the black hole spin. The gas of the disk moves on nearly geodesic equatorial circular orbits. Imposing the conservation of mass, energy, and angular momentum, we can derive the time-averaged radial structure of the disk. The time-averaged energy flux 
emitted from the surface of the disk turns out to be~\citep{nt2}
\be\label{eq-fff}
\mathcal{F}(r) = \frac{\dot{M}}{4 \pi M^2} F(r) \, ,
\ee
where $F(r)$ is the dimensionless function
\be\label{eq-f}
F(r) = - \frac{\partial_r \Omega}{(E - \Omega L_z)^2} 
\frac{M^2}{\sqrt{-G}}
\int_{r_{\rm in}}^{r} (E - \Omega L_z) 
(\partial_\rho L_z) \, d\rho \, .
\nonumber\\
\ee
$E$, $L_z$, and $\Omega$ are, respectively, the conserved specific energy, the conserved axial-component of the specific angular momentum, and the angular velocity for equatorial circular geodesics (for their derivation, see e.g. Ref.~\cite{nk4})
\be
E &=& - \frac{g_{tt} + \Omega g_{t\phi}}{\sqrt{- g_{tt} 
- 2 \Omega g_{t\phi} - \Omega^2 g_{\phi\phi}}}\, , \\
L_z &=& \frac{g_{t\phi} + \Omega g_{\phi\phi}}{\sqrt{- g_{tt} 
- 2 \Omega g_{t\phi} - \Omega^2 g_{\phi\phi}}} \, , \\
\Omega &=& \frac{- \partial_r g_{t\phi} \pm \sqrt{\left(\partial_r g_{t\phi}\right)^2 - 
\left(\partial_r g_{tt}\right) \left(\partial_r g_{\phi\phi}\right)}}{\partial_r g_{\phi\phi}} \, ,
\ee
where the upper (lower) sign in the expression of $\Omega$ refers to corotating (counterrotating) orbits, namely orbits with angular momentum parallel (antiparallel) to the black hole spin, and the metric $g_{\mu\nu}$ is written in the canonical form for stationary and axisymmetric spacetimes; that is, the line element is
\be
ds^2 = g_{tt} dt^2 + 2 g_{t\phi} dt d\phi + g_{rr} dr^2 
+ g_{\theta\theta} d\theta^2 + g_{\phi\phi} d\phi^2 \, . \nonumber\\
\ee
In Eq.~(\ref{eq-f}), $G = - \alpha^2 g_{rr} g_{\phi\phi}$ is the determinant of the near equatorial plane metric, where $\alpha^2 = g_{t\phi}^2/g_{\phi\phi} - g_{tt}$ is the lapse function, and $r_{\rm in}$ is the inner radius of the accretion disk. In general, it is assumed that $r_{\rm in}$ corresponds to the radius of the innermost stable circular orbit (ISCO), $r_{\rm ISCO}$. However, in principle it is possible to have a truncated disk with $r_{\rm in} > r_{\rm ISCO}$.

The accretion disk is assumed to be locally in thermal equilibrium and at every point emits like a blackbody. Since the system is axisymmetric, we can define an effective temperature at every radius $T_{\rm eff} (r)$ from the relation $\mathcal{F}(r) = \sigma T^4_{\rm eff}$, where $\sigma$ is the Stefan-Boltzmann constant. Non-thermal effects are non-negligible when the disk temperature is very high; this, for example, is the case of the inner edge of disks around stellar mass black holes, where $T_{\rm eff}$ can be up to $\sim 10^7$~K. Non-thermal effects, largely due to electron scattering in the disk atmosphere, can be taken into account by introducing the color factor (or hardening factor) $f_{\rm col}$. The color temperature is defined as $T_{\rm col} (r) = f_{\rm col} T_{\rm eff}$. The local specific intensity of the radiation emitted by the disk (measured, for instance, in erg~s$^{-1}$~cm$^{-2}$~str$^{-1}$~Hz$^{-1}$) turns out to be (reintroducing the speed of light $c$)
\be\label{eq-i-bb}
I_{\rm e}(\nu_{\rm e}) = \frac{2 h \nu^3_{\rm e}}{c^2} \frac{1}{f_{\rm col}^4} 
\frac{\Upsilon}{\exp\left(\frac{h \nu_{\rm e}}{k_{\rm B} T_{\rm col}}\right) - 1} \, ,
\ee
where $\nu_{\rm e}$ is the photon frequency in the rest-frame of the gas, $h$ is Planck's constant, and $k_{\rm B}$ is the Boltzmann constant. Here $\Upsilon$ is a function of the angle between the propagation direction of the photon emitted by the disk and the normal to the disk surface, say, $\xi$. The two most common choices are $\Upsilon = 1$
(isotropic emission) and $\Upsilon = \frac{1}{2} + \frac{3}{4} \cos\xi$ (limb-darkened emission).

The flux of the disk (measured, for instance, in erg~s$^{-1}$~cm$^{-2}$~Hz$^{-1}$) at the photon frequency $\nu_{\rm o}$ measured by a distant observer can be written as
\be\label{eq-thin-Fobs}
F_{\rm o} (\nu_{\rm o}) &=& \int I_{\rm o}(\nu_{\rm o}, X, Y) d\tilde{\Omega} \nonumber\\
&=& \int g^3 I_{\rm e}(\nu_{\rm e}, r_{\rm e}, \vartheta_{\rm e}) d\tilde{\Omega} \, ,
\ee
where $I_{\rm o}$ is the specific intensity of the radiation detected by the distant observer, $X$ and $Y$ are the Cartesian coordinates of the image of the disk in the plane of the distant observer, $d\tilde{\Omega} = dX dY/D^2$ is the element of the solid angle subtended by the image of the disk in the observer's sky, $D$ is the distance of the observer from the source, $r_{\rm e}$ is the emission radius in the disk and $\vartheta_{\rm e}$ is the emission angle (which can be different from the angle between the black hole spin and the line of sight of the observer because of the effect of light bending), and $g$ is the redshift factor
\be\label{eq-red}
g = \frac{\nu_{\rm o}}{\nu_{\rm e}} = 
\frac{k_\alpha u^{\alpha}_{\rm o}}{k_\beta u^{\beta}_{\rm e}}\, .
\ee
$k^\alpha$ is the photon 4-momentum, $u^{\alpha}_{\rm o} = (1,0,0,0)$ is the 4-velocity of the distant observer, and $u^{\alpha}_{\rm e} = (u^t_{\rm e},0,0, \Omega u^t_{\rm e})$ is the 4-velocity of the emitter. In Eq.~(\ref{eq-thin-Fobs}), we exploit the relation $I_{\rm e}(\nu_{\rm e})/\nu_{\rm e}^3 = I_{\rm o} (\nu_{\rm o})/\nu_{\rm o}^3$ following from Liouville's theorem.

From the normalization condition $g_{\mu\nu}u^{\mu}_{\rm e}u^{\nu}_{\rm e} = -1$, we can write the temporal component of the 4-velocity of the emitter as
\be
u^t_{\rm e} = - \frac{1}{\sqrt{-g_{tt} - 2 g_{t\phi} \Omega - g_{\phi\phi} \Omega^2}} \, ,
\ee
and then the redshift factor $g$ is
\be\label{eq-red-g}
g = \frac{\sqrt{-g_{tt} - 2 g_{t\phi} \Omega - g_{\phi\phi} \Omega^2}}{1 + 
\lambda \Omega} \, .
\ee
$\lambda = k_\phi/k_t$ is a constant of motion along the photon path.

Deviations from the Novikov-Thorne model in real astrophysical disks inevitably introduce systematic uncertainties in the estimates of the model parameters, which can limit our capability of using this method to test the Kerr hypothesis. This issue was studied in Refs.~\cite{2-cfm1,2-cfm2}, where it was shown that measurements using the continuum-fitting method are currently limited by the uncertainties on the black hole mass, black hole distance, and inclination angle of the accretion disk, while deviations from Novikov-Thorne disks are subdominant.

%%%%%%%%%%%%%%%%%%%%%%%%%%%%%%%

\section{Transfer function \label{s-trans}}

Ray-tracing calculations to evaluate the spectrum of the disk are too time consuming to be performed during the data analysis stage. The formalism of the transfer function proposed by Cunningham~\cite{cun} permits one to store all of the necessary details about the spacetime metric in a FITS file. The latter can be called during the data analysis phase.

The expression of the observed flux in Eq.~(\ref{eq-thin-Fobs}) can be recast in the following form
\be\label{eq-Fobs}
F_{\rm o} (\nu_{\rm o}) 
= \frac{1}{D^2} \int_{r_{\rm in}}^{r_{\rm out}} \int_0^1&
\pi r_{\rm e} \frac{ g^2}{\sqrt{g^* (1 - g^*)}} f(g^*,r_{\rm e},i)
\nonumber\\ & \times \,
I_{\rm e}(\nu_{\rm e},r_{\rm e},\vartheta_{\rm e}) \, dg^* \, dr_{\rm e} \, .
\ee
Here $r_{\rm in}$ and $r_{\rm out}$ are, respectively, the inner and outer radius of the accretion disk, and $f$ is the transfer function~\cite{cun}
\be\label{eq-trf}
f(g^*,r_{\rm e},i) = \frac{1}{\pi r_{\rm e}} g 
\sqrt{g^* (1 - g^*)} \left| \frac{\partial \left(X,Y\right)}{\partial \left(g^*,r_{\rm e}\right)} \right| \, ,
\ee
where the relative redshift factor $g^*$ is defined as
\be
g^* = \frac{g - g_{\rm min}}{g_{\rm max} - g_{\rm min}} \, ,
\ee
and ranges from 0 to 1. $g_{\rm max}=g_{\rm max}(r_{\rm e},i)$ and $g_{\rm min}=g_{\rm min}(r_{\rm e},i)$ are, respectively, the maximum and minimum values of the redshift factor $g$ for the photons emitted from the radial coordinate $r_{\rm e}$ and detected by a distant observer with polar coordinate $i$. $\left| \partial \left(X,Y\right)/\partial \left(g^*,r_{\rm e}\right) \right|$ indicates the Jacobian. The transfer function acts as an integration kernel, permitting us to calculate the observed spectrum from the local spectrum at any point of the disk. Note that the specific intensity $I_{\rm e}$, $\nu_{\rm e}$, and $\vartheta_{\rm e}$ must be written in terms of $g^*$ and $r_{\rm e}$. In our model, we ignore secondary and higher order images of the disk generated by photons crossing the equatorial plane between the black hole and the inner edge of the disk.

The transfer function $f (g^*, r_{\rm e}, i)$ is completely determined by the spacetime metric and the viewing angle of the distant observer, $i$. For a specific value of the emission radius $r_{\rm e}$ and of the viewing angle $i$, the transfer function is a closed curve parametrized by $g^*$. This is true except in the special cases $i = 0$ and $\pi/2$, where the transfer function is not well defined. For every emission radius $r_{\rm e}$, there is only one point in the disk for which $g^* = 1$ and only one point for which $g^* = 0$. These points are connected by two curves, so we have two branches of the transfer function, say $f^{(1)} (g^*,r_{\rm e},i)$ and $f^{(2)}(g^*,r_{\rm e},i)$. In the case of isotropic emission ($I_{\rm e}$ independent of $\vartheta_{\rm e}$ and of the emission azimuthal angle) in an axisymmetric system (e.g. no orbiting spots), Eq.~(\ref{eq-Fobs}) reduces to
\begin{widetext}
\be
F_{\rm o} (\nu_{\rm o}) 
&=& \frac{1}{D^2} \int_{r_{\rm in}}^{r_{\rm out}} \int_0^1
\frac{\pi r_{\rm e} \, g^2}{\sqrt{g^* (1 - g^*)}} 
\left[ f^{(1)} (g^*,r_{\rm e},i) + f^{(2)} (g^*,r_{\rm e},i) \right] 
I_{\rm e}(\nu_{\rm e},r_{\rm e}) \, dg^* \, dr_{\rm e} \, . \nonumber\\
\ee
If $I_{\rm e}$ depends on $\vartheta_{\rm e}$, it is necessary to perform the integral twice, one for the upper branch and one for the lower one, and Eq.~(\ref{eq-Fobs}) becomes
\be\label{eq-trf-1-2-b}
\hspace{-1.2cm}
F_{\rm o} (\nu_{\rm o}) 
&=& \frac{1}{D^2} \int_{r_{\rm in}}^{r_{\rm out}} \int_0^1
\frac{\pi r_{\rm e} \, g^2}{\sqrt{g^* (1 - g^*)}} 
f^{(1)} (g^*,r_{\rm e},i)
I_{\rm e}(\nu_{\rm e},r_{\rm e},\vartheta_{\rm e}^{(1)}) \, dg^* \, dr_{\rm e} \nonumber\\
&& + \frac{1}{D^2} \int_{r_{\rm in}}^{r_{\rm out}} \int_0^1
\frac{\pi r_{\rm e} \, g^2}{\sqrt{g^* (1 - g^*)}} 
f^{(2)} (g^*,r_{\rm e},i)
I_{\rm e}(\nu_{\rm e},r_{\rm e},\vartheta_{\rm e}^{(2)}) \, dg^* \, dr_{\rm e} \, ,
\ee
\end{widetext}
where $\vartheta_{\rm e}^{(1)}$ and $\vartheta_{\rm e}^{(2)}$ indicate the emission angles with relative redshift factor $g^*$, respectively in branch~1 and branch~2.

The Jacobian in the expression of the transfer function is calculated from
\begin{align}\label{eq:jacob}
\left| \frac{\partial \left(X,Y\right)}{\partial \left(g^*,r_{\rm e}\right)} \right|
= \left| 
\frac{\partial r_{\rm e}}{\partial X}\frac{\partial g^*}{\partial Y}
- \frac{\partial r_{\rm e}}{\partial Y}\frac{\partial g^*}{\partial X} \right| ^{-1} \, .
\end{align}

For the calculation of the transfer function, we employ the code already described in~\cite{noi1,noi2}. However, here we use a different sampling because the thermal component decreases more slowly as the radius increases than the reflection component and it is thus necessary to sample even at larger radii. The value of the transfer function is calculated for 200~radii, ranging from the ISCO to $10^6$~$M$. At every radius, the transfer function is evaluated at 47~different values of $g^*$, which are not evenly distributed from 0 to 1 but their density is higher near 0 and 1 and decreases moving to 0.5.

%%%%%%%%%%%%%%%%%%%%%%%%%%%%%%%

\section{The XSPEC model {\sc nkbb} \label{s-xspec}}

As we did in Refs.~\cite{noi1,noi2}, we construct our model for the Johannsen metric~\cite{tj}, but we can easily implement any other stationary, axisymmetirc, and asymptotically flat black hole spacetime without pathological properties. In Boyer-Lindquist-like coordinates, the line element reads~\cite{tj}
\be\label{eq-jm}
ds^2 &=&-\frac{\tilde{\Sigma}\left(\Delta-a^2A_2^2\sin^2\theta\right)}{B^2}dt^2
+\frac{\tilde{\Sigma}}{\Delta A_5}dr^2+\tilde{\Sigma} d\theta^2 \nonumber\\
&&-\frac{2a\left[\left(r^2+a^2\right)A_1A_2-\Delta\right]\tilde{\Sigma}\sin^2\theta}{B^2}dtd\phi \nonumber\\
&&+\frac{\left[\left(r^2+a^2\right)^2A_1^2-a^2\Delta\sin^2\theta\right]\tilde{\Sigma}\sin^2\theta}{B^2}d\phi^2 \, ,
\ee
where $M$ is the black hole mass, $a = J/M$, $J$ is the black hole spin angular momentum, $\tilde{\Sigma} = \Sigma + f$, and
\be
&&  \Sigma = r^2 + a^2 \cos^2\theta \, , \qquad
\Delta = r^2 - 2 M r + a^2 \, , \nonumber\\
&& B = \left(r^2+a^2\right)A_1-a^2A_2\sin^2\theta \, .
\ee
The functions $f$, $A_1$, $A_2$, and $A_5$ are defined as
\be
f &=& \sum^\infty_{n=3} \epsilon_n \frac{M^n}{r^{n-2}} \, , \\
A_1 &=& 1 + \sum^\infty_{n=3} \alpha_{1n} \left(\frac{M}{r}\right)^n \, , \\
A_2 &=& 1 + \sum^\infty_{n=2} \alpha_{2n} \left(\frac{M}{r}\right)^n \, , \\
A_5 &=& 1 + \sum^\infty_{n=2} \alpha_{5n} \left(\frac{M}{r}\right)^n \, ,
\ee
where $\{ \epsilon_n \}$, $\{ \alpha_{1n} \}$, $\{ \alpha_{2n} \}$, and $\{ \alpha_{5n} \}$ are four infinite sets of deformation parameters and the Kerr metric is recovered when all deformation parameters vanish. Note that this form of the metric already recovers the correct Newtonian limit and passes Solar System experiments without fine-tuning.

The deformation parameters alter the geometry of the spacetime and, consequently, the motion of massive and massless particles. Every set of deformation parameters has its own effect on the spacetime, while within the same set the difference among different deformation parameters is only on their radial profile. The deformation parameter $\alpha_{13}$ is that with the strongest impact on the structure of the accretion disk and thus on its thermal spectrum. For $\alpha_{13} > 0$, the gravitational force gets stronger and thus the value of the ISCO radius increases. For $\alpha_{13} < 0$, we have the opposite effect; that is, the gravitational force gets weaker and the value of the ISCO radius decreases. In what follows, we will only focus on this deformation parameter, setting all others to zero, but the implementation of our model with other non-vanishing deformation parameters is straightforward.

The Johannsen spacetime can present some pathological properties (spacetime singularities, regions with closed time-like curves, etc.) for certain values of the spin and the deformation parameters. We thus impose some restrictions on the spin parameter $a_* = a/M = J/M^2$ and of the deformation parameter $\alpha_{13}$. The condition for $a_*$ is
\be
- 1 < a_* < 1 \, .
\ee 
As in the Kerr metric, this is simply the condition for the existence of the event horizon: for $| a_* | > 1$, there is no horizon and the central singularity is naked. The condition for $\alpha_{13}$ is~\cite{tj,c2}
\be
\label{eq-constraints}
\alpha_{13} > -\frac{1}{2}\left(1+\sqrt{1-a^2_*}\right)^{4} \, .
\ee

\begin{figure}[b]
\begin{center}
\includegraphics[width=9.0cm]{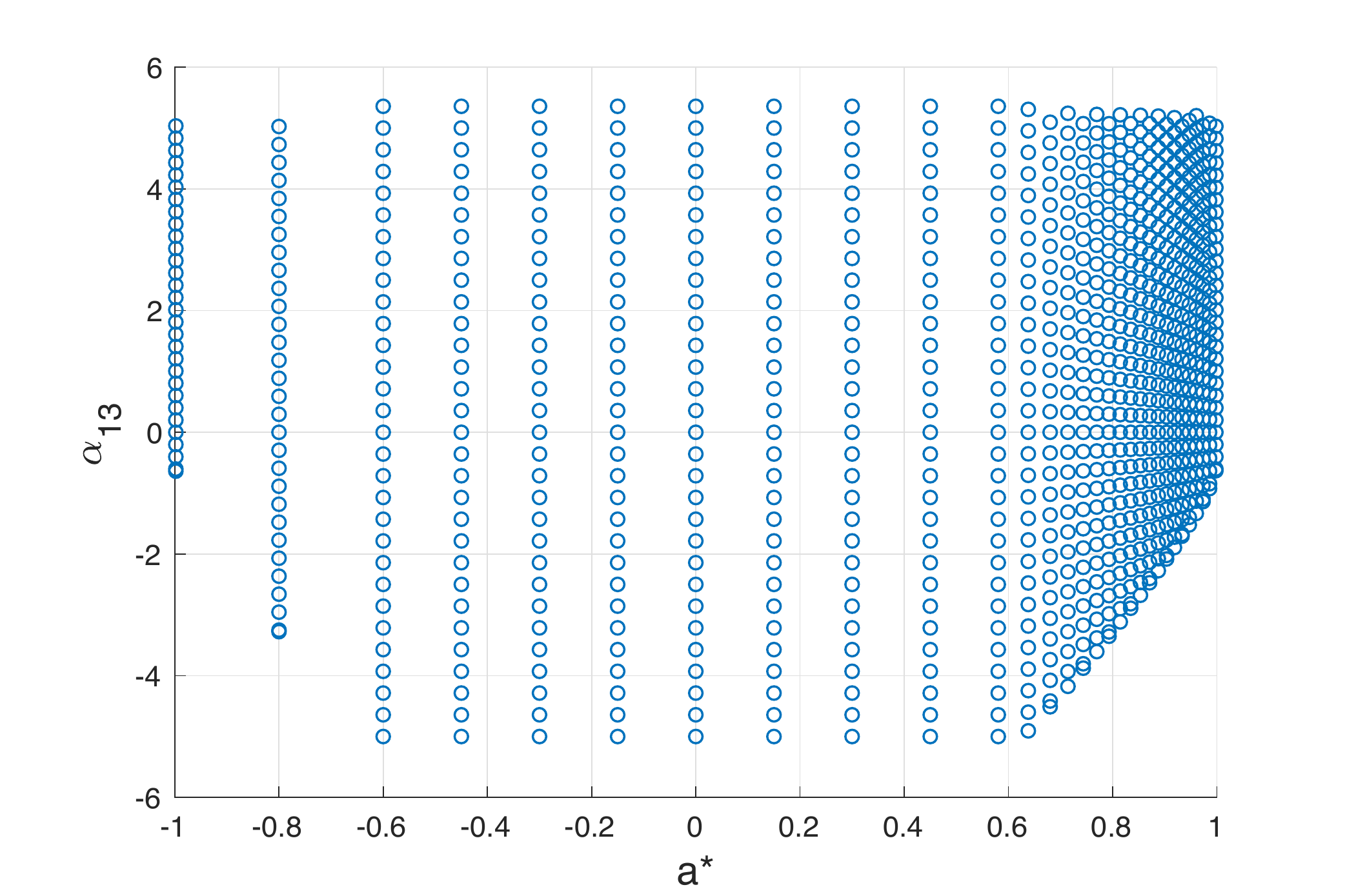}
\end{center}
\vspace{-0.4cm}
\caption{Grid points in the FITS file for the spin parameter $a_*$ and the deformation parameter $\alpha_{13}$. \label{f-grid}}
\end{figure}

The transfer function is evaluated for a grid of points in the 3D parameter space $( a_*, \alpha_{13}, i )$ and stored in a FITS file. The grid points for the viewing angle $i$ are distributed evenly in $0 < \cos i < 1$. The distribution of grid points in the remaining 2D parameter space $( a_*, \alpha_{13})$ is shown in Fig.~\ref{f-grid}. Note that the density of points increases as we move from $a_* = -1$ to $a_* = 1$ because the ISCO radius changes faster for fast-rotating black holes and co-rotating disks and we thus need to better sample the parameter space.

\begin{figure*}[t]
\begin{center}
\includegraphics[width=8.5cm]{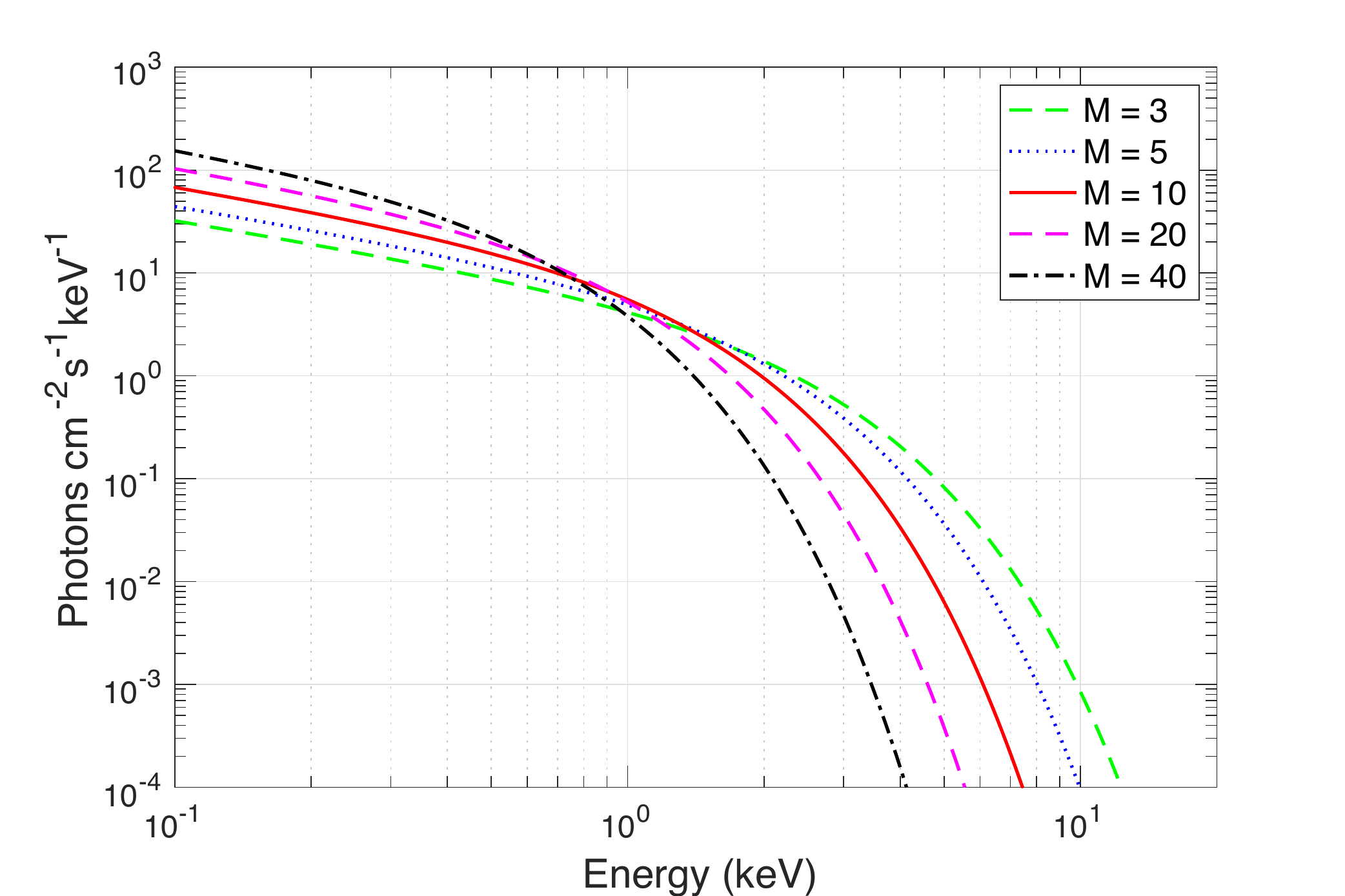}
\includegraphics[width=8.5cm]{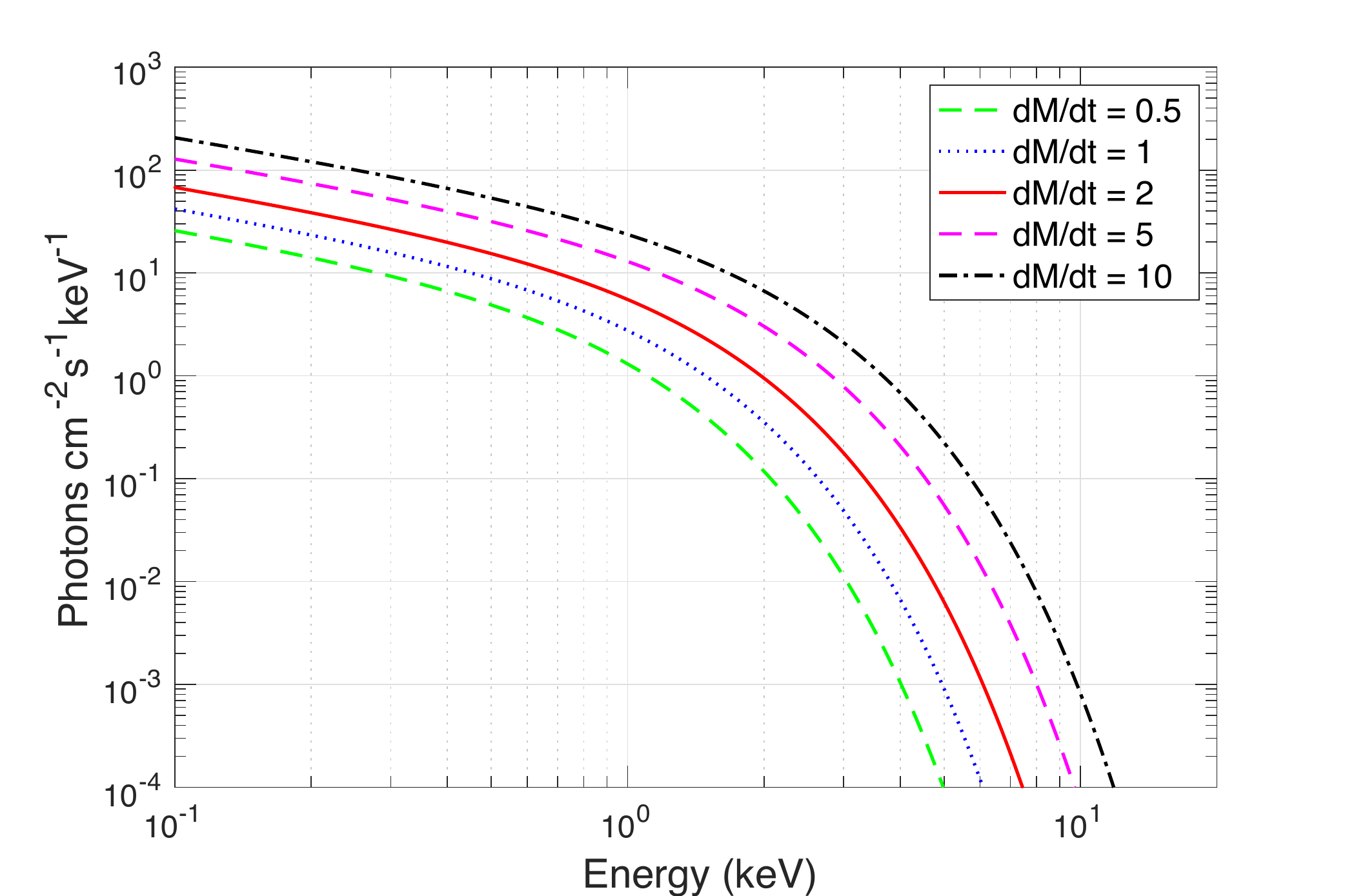} \\ \vspace{0.2cm}
\includegraphics[width=8.5cm]{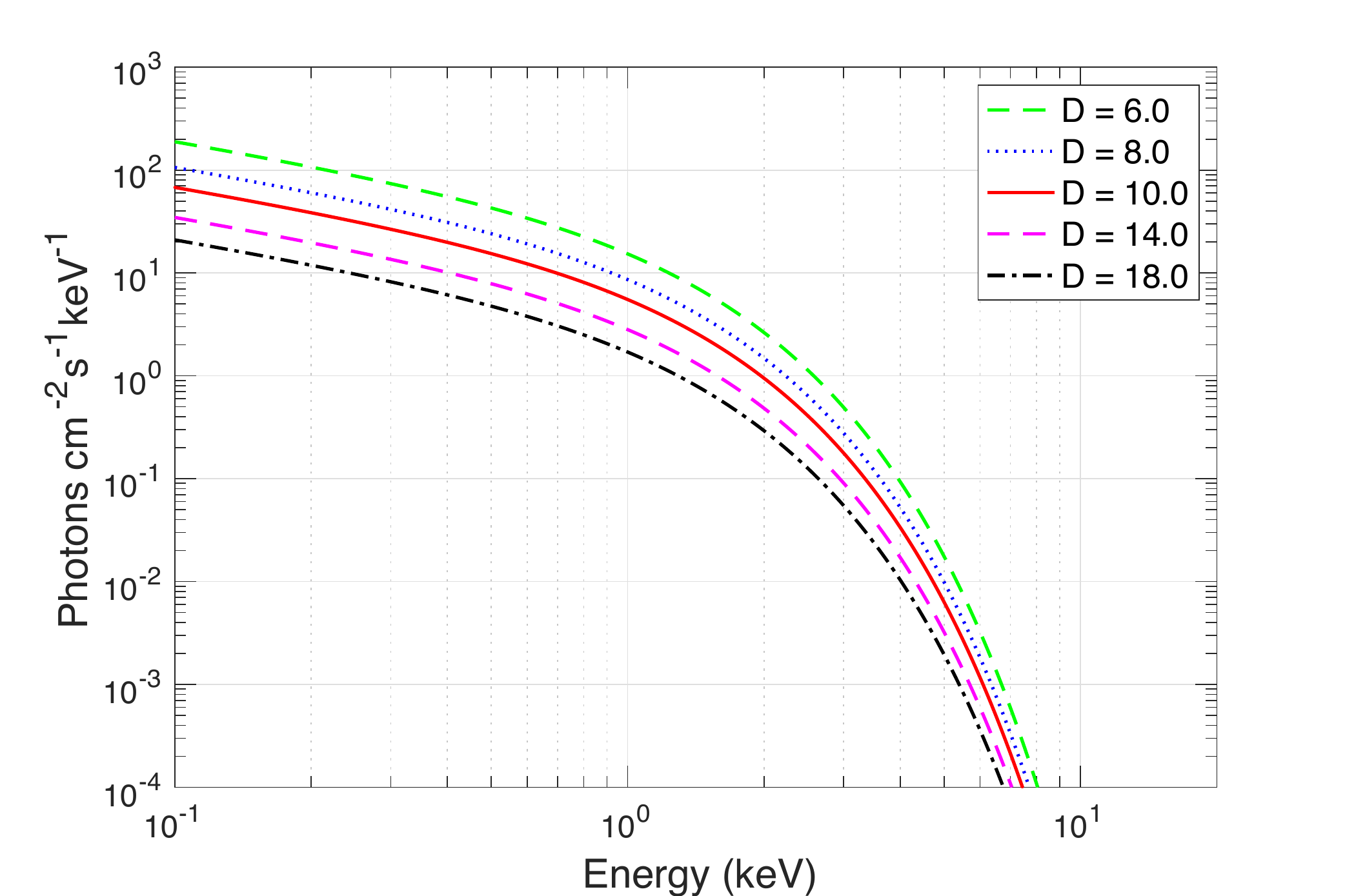}
\includegraphics[width=8.5cm]{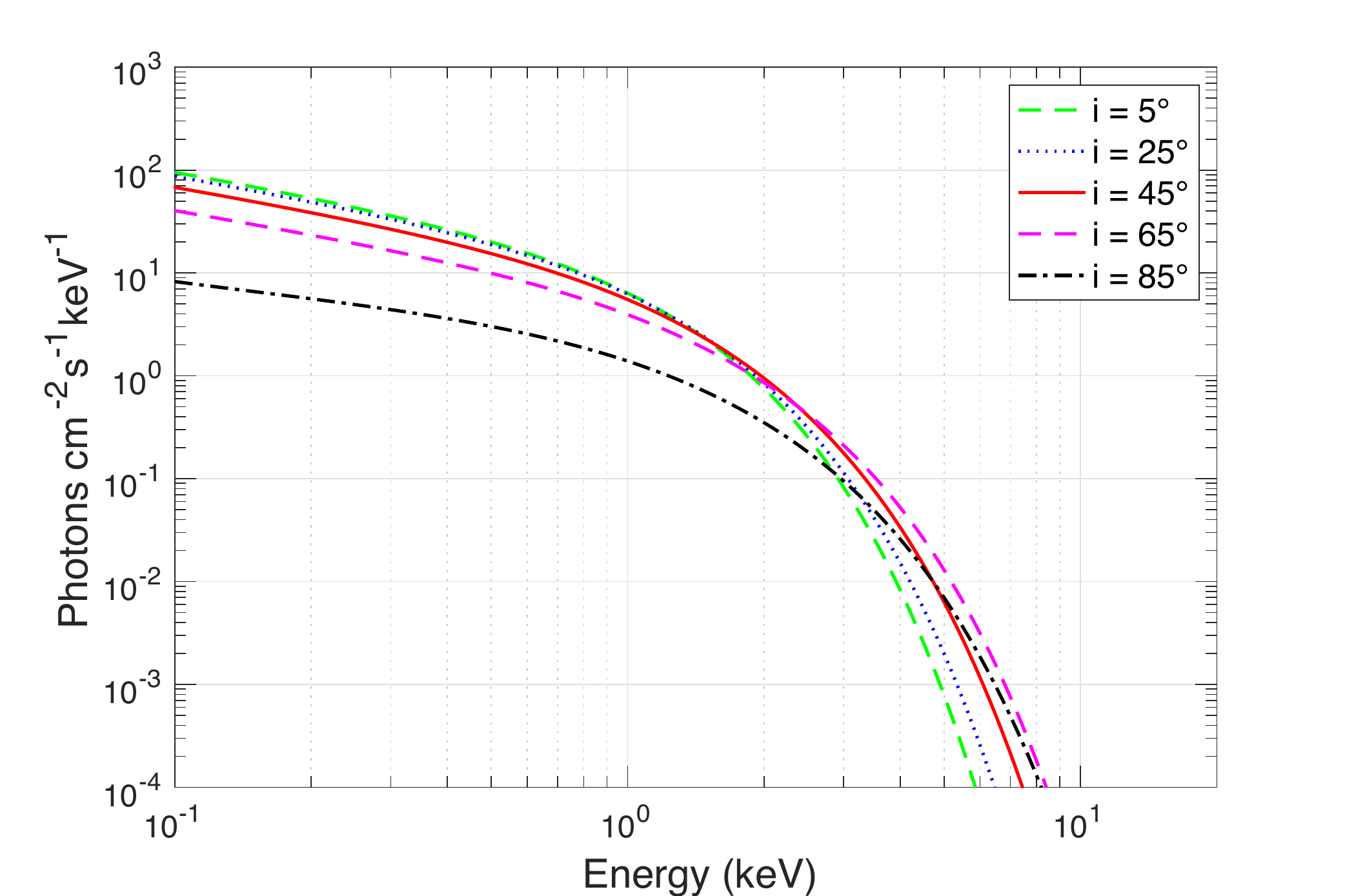} \\ \vspace{0.2cm}
\includegraphics[width=8.5cm]{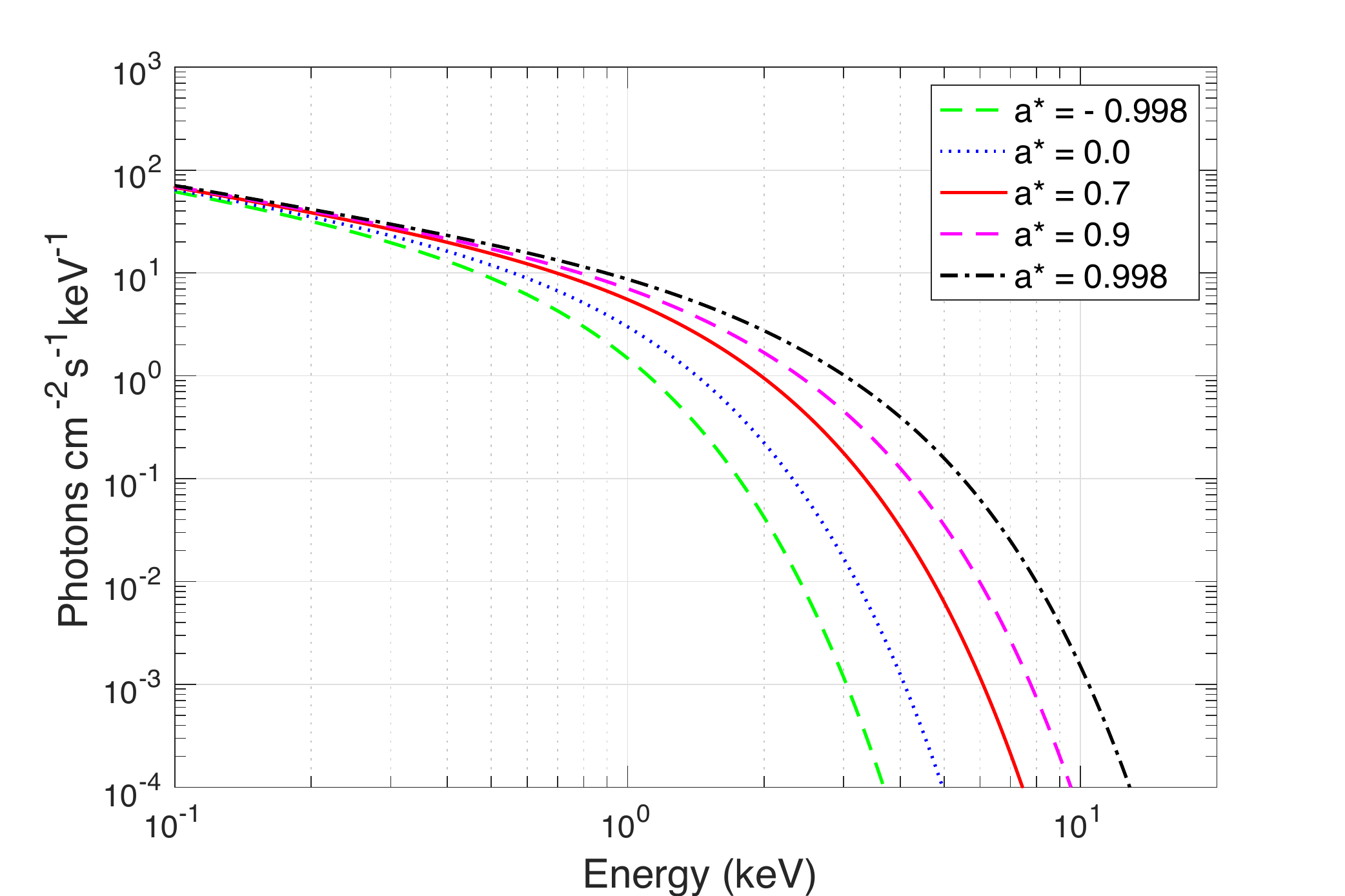}
\includegraphics[width=8.5cm]{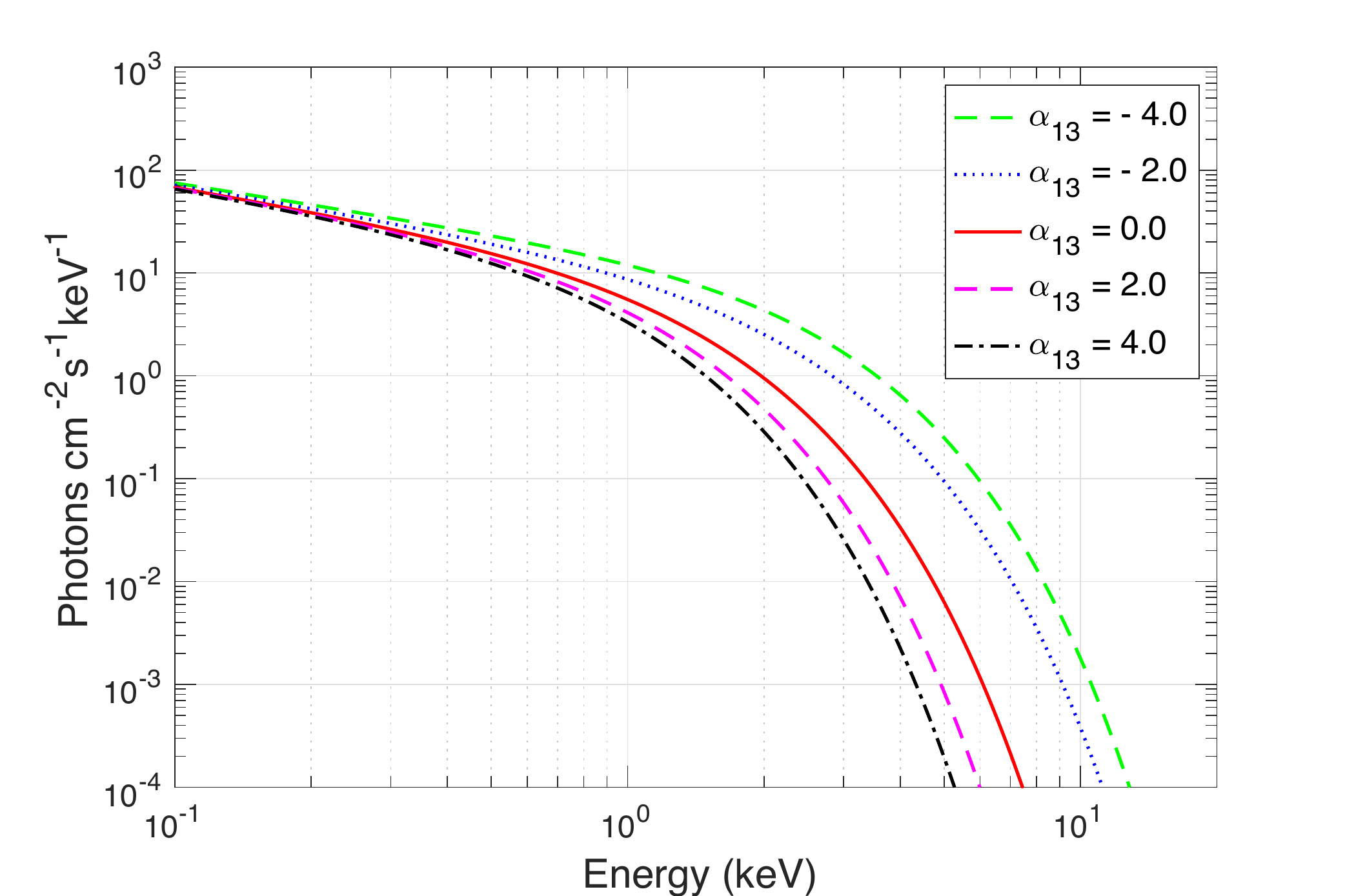}
\end{center}
\vspace{-0.2cm}
\caption{Thermal spectra of accretion disks as calculated by {\sc nkbb} for different values of the model parameters. $M$ in $M_\odot$, $\dot{M}$ in $10^{18}$~g/s, and $D$ in kpc. When not shown, the values of the model parameters are: $M = 10 \, M_\odot$, $\dot{M} = 2 \cdot 10^{18}$~g/s, $D = 10$~kpc, $i = 45^\circ$, $a_* = 0.7$, $\alpha_{13} = 0$. In these simulations we have assumed $f_{\rm col} = \Upsilon = 1$. \label{f-para}}
\end{figure*}

In the end, {\sc nkbb} has 8~parameters: the black hole mass $M$, the black hole distance $D$, the mass accretion rate $\dot{M}$ (which becomes the effective mass accretion rate in the presence of a non-vanishing torque at the inner edge of the disk~\cite{kerrbb}), the viewing angle of the observer $i$, the black hole spin $a_*$, the deformation parameter $\alpha_{13}$, the color factor $f_{\rm col}$, and the function $\Upsilon$. Synthetic thermal spectra calculated by {\sc nkbb} for different values of the model parameters are shown in Fig.~\ref{f-para}. The mass $M$ changes the temperature of the disk, as follows from the relation $\mathcal{F}(r) = \sigma T^4_{\rm eff}$ and Eq.~(\ref{eq-fff}), as well as the size of the disk: if $M$ decreases/increases, the spectrum gets harder/softer. The mass accretion rate $\dot{M}$ only changes the temperature of the disk through Eq.~(\ref{eq-fff}). The distance of the source $D$ simply changes the normalization of the spectrum. The inclination angle $i$ regulates the Doppler boosting: for higher/lower $i$, the effect of the Doppler boosting increases/decreases and (neglecting the effect of light bending) the area of the observed emitting region decreases/increases. Last, the spin $a_*$ and the deformation parameter $\alpha_{13}$ have rather similar effects and mainly change the position of the ISCO radius: if $a_*$ increases and/or $\alpha_{13}$ decreases, the ISCO radius decreases and the disk gets a higher temperature area emitting thermal photons, so the spectrum gets harder. As we can qualitatively understand from these plots, the estimates of the parameters $a_*$ and $\alpha_{13}$ will be quite correlated, as the impact of these parameters on the thermal spectrum is similar. The impact of all these parameters on the thermal spectra of thin disks has already been discussed in the literature; see, for instance, Refs.~\cite{nk4,nk6}.

%%%%%%%%%%%%%%%%%%%%%%%%%%%%%%%

\section{Comparison with existing models \label{s-comp}}

The XSPEC model commonly used to describe the thermal spectrum of the accretion disk around a Kerr black hole is {\sc kerrbb}~\cite{kerrbb}. We can test the accuracy of {\sc nkbb} by comparing the two models in the Kerr spacetime; that is, we set $\alpha_{13} = 0$ in {\sc nkbb}. Such a comparison can give us a rough estimate of the accuracy of our model.

\begin{figure*}[t]
\begin{center}
\includegraphics[width=8.5cm]{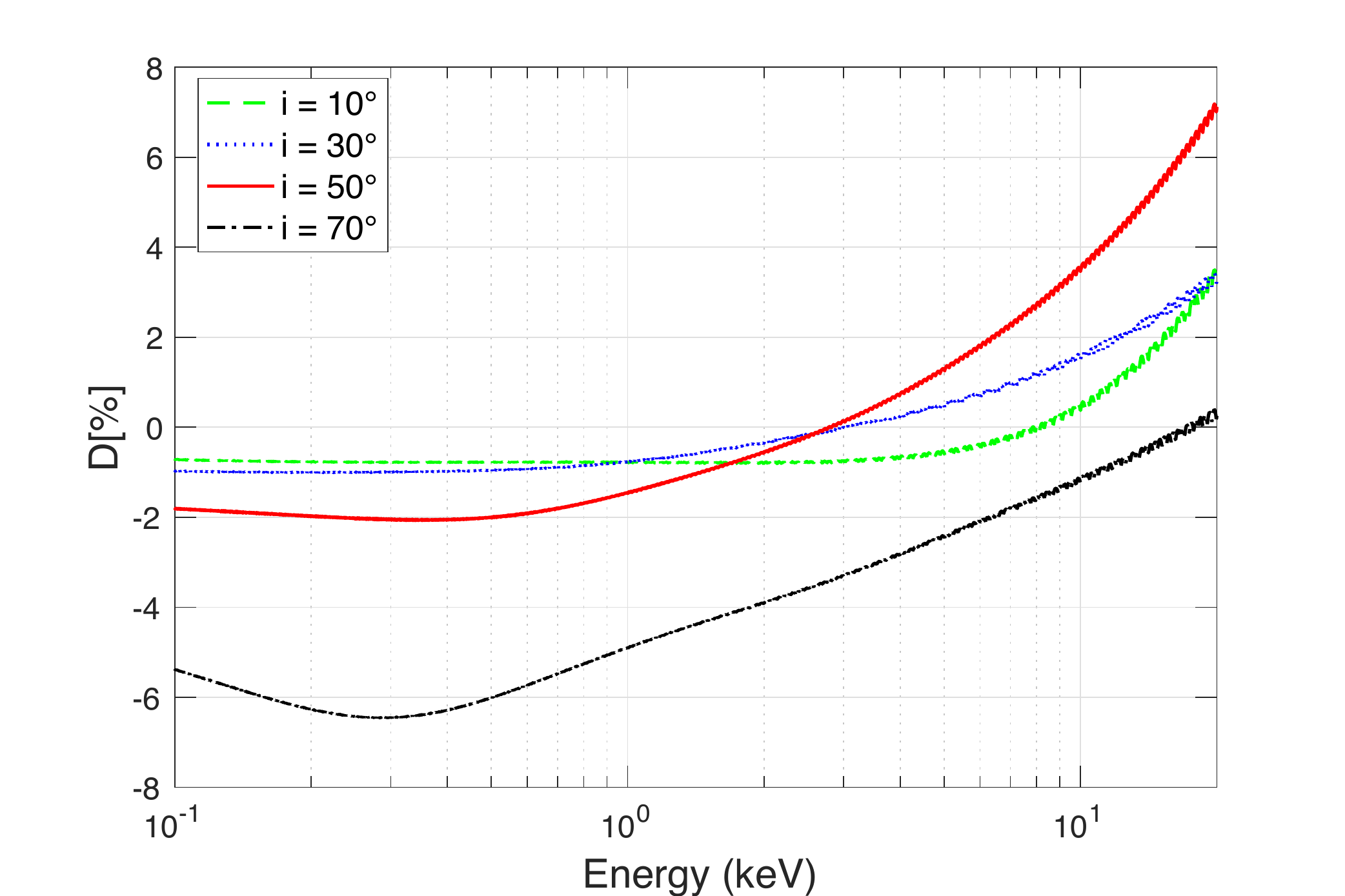}
\includegraphics[width=8.5cm]{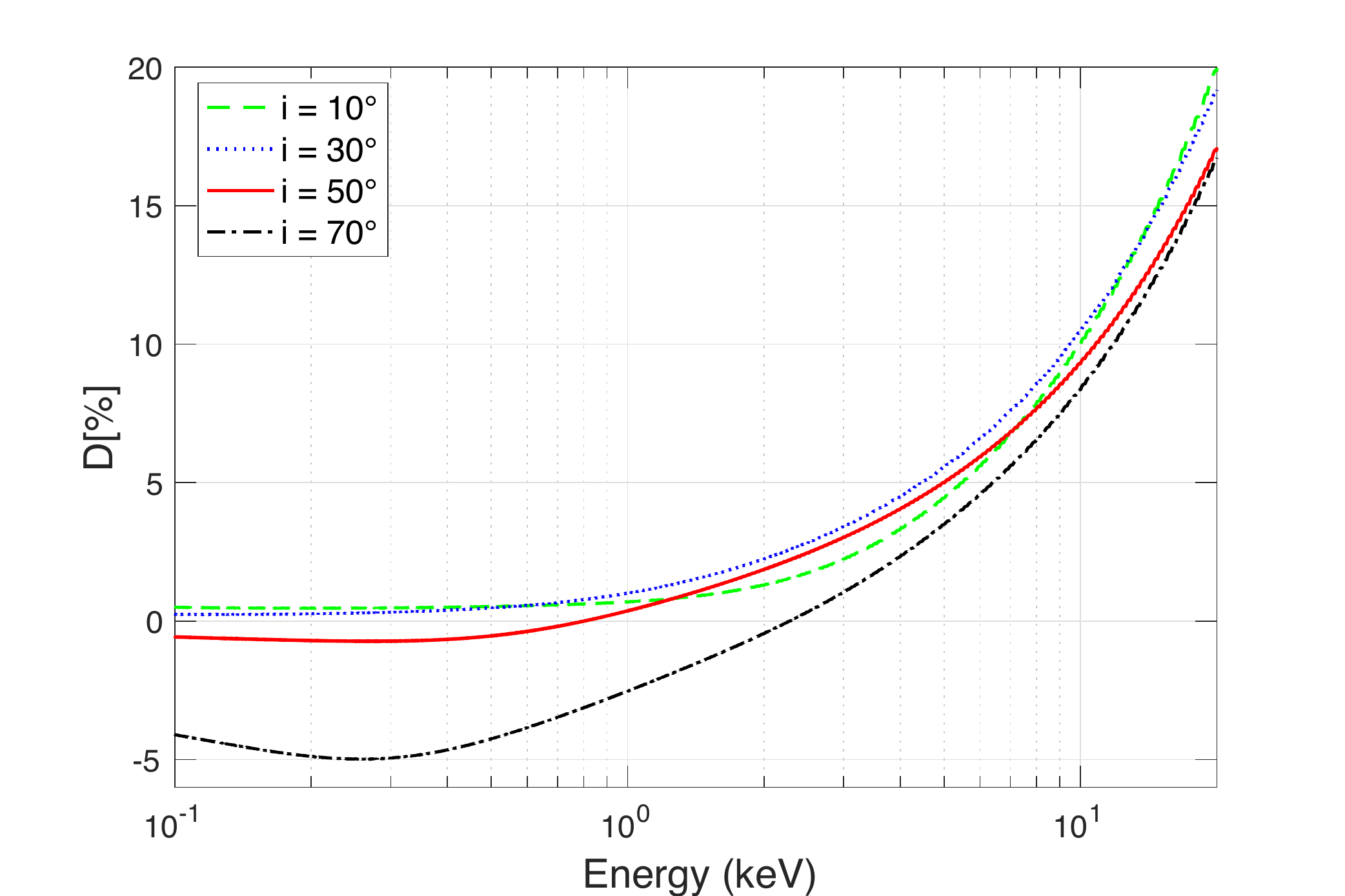} \\ \vspace{0.2cm}
\includegraphics[width=8.5cm]{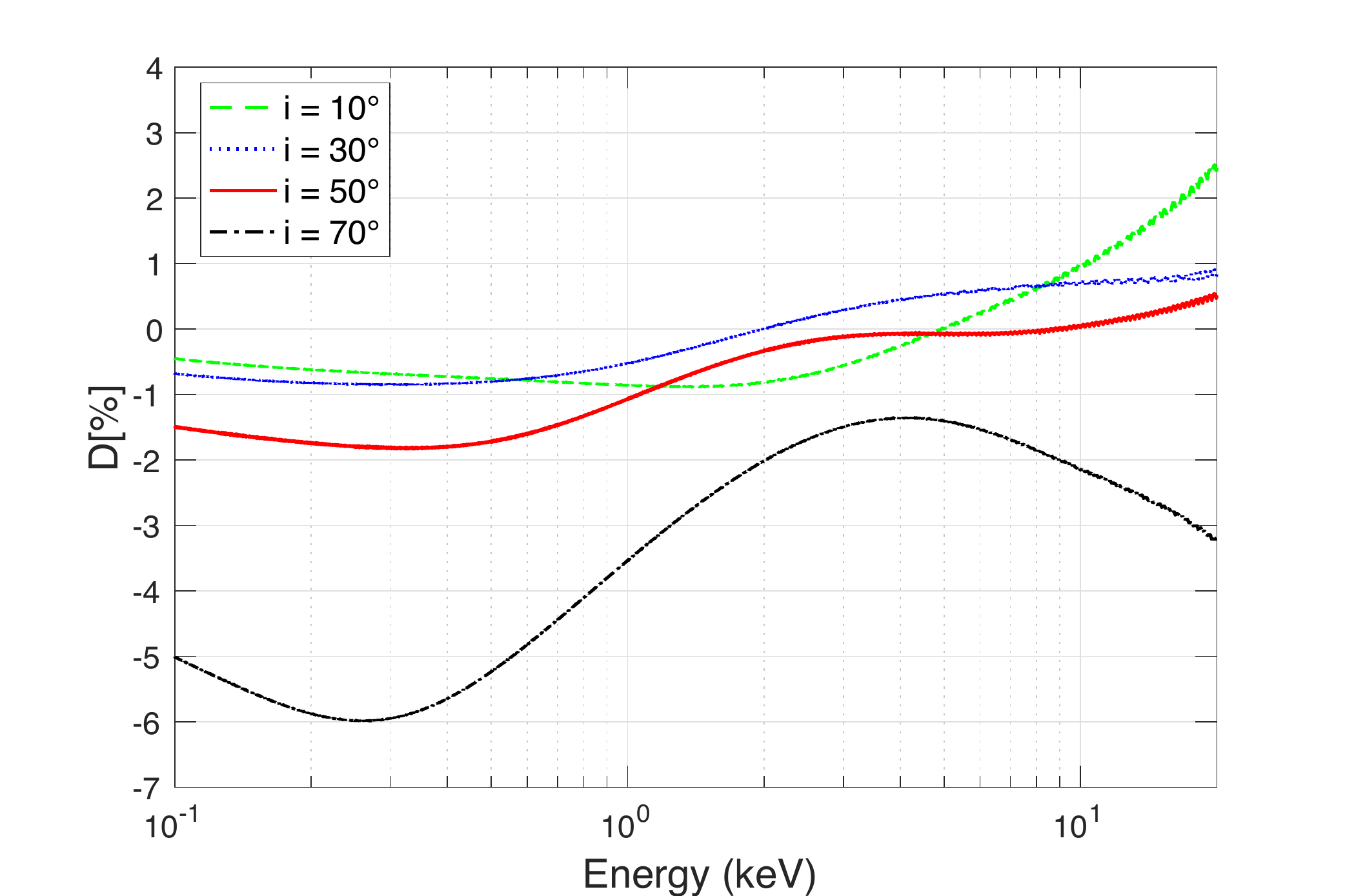}
\end{center}
\vspace{-0.2cm}
\caption{Discrepancies between the spectra calculated by {\sc nkbb} and {\sc kerrbb} in the Kerr spacetime. The spin parameters is $a_* = -0.5$ (top left panel), 0.5 (top right panel), and 0.998 (bottom panel) and the inclination angles are $i = 10^\circ$, $30^\circ$, $50^\circ$, and $70^\circ$. The black hole mass is fixed at $10 \, M_\odot$, the mass accretion rate is fixed at $10^{19}$~g/s, and $f_{\rm col}$ is fixed at 1. \label{f-dis}}
\end{figure*}

Fig.~\ref{f-dis} shows the differences between spectra calculated by our model and by {\sc kerrbb} for $a_* = -0.5$ (top left panel), 0.5 (top right panel), and 0.998 (bottom panel) when the inclination angle of the disk with respect to the line of sight of the distant observer is $i = 10^\circ$, $30^\circ$, $50^\circ$, and $70^\circ$. These 12~cases are used to check if there is an problem in the construction of our FITS file of the transfer functions and in the calculations of thermal spectra from interpolation of the grid points in the FITS file. The black hole mass $M$ is set to 10~$M_\odot$, the mass accretion rate $\dot{M}$ to $10^{19}$~g/s, and we assume $f_{\rm col} = \Upsilon = 1$. We show the relative differences between the two models and not the spectra because the latter would perfectly overlap in a $\log$-$\log$ plot and would not permit us to see the discrepancy between the calculations of {\sc kerrbb} and {\sc nkbb}. For $i = 10^\circ$, $30^\circ$, and $50^\circ$, the agreement between the two models is within a few percent for energies below 7-8~keV, which is the range relevant for the continuum-fitting method. At higher energies, we see that the discrepancies between the two models increase. However, the accuracy at high energies is not so important because the luminosity falls off and the thermal component from the disk is normally negligible with respect to a power-law component originated from inverse Compton scattering in the hot gas around the black hole (moreover, for $f_{\rm col} \approx 1.5-1.7$ the discrepancy starts to increase at even higher energies). For the case $i = 70^\circ$, the difference between the two models is already around 5\% at low energies. The problem here is in our model, which has a lower accuracy for $i \gtrsim 70^\circ$. We do not know the exact origin of the problem and we hope to fix it in the next version of the model.

%%%%%%%%%%%%%%%%%%%%%%%%%%%%%%%

\section{Simulations \label{s-sim}}

If we assume that the spacetime metric around black holes is described by the Kerr solution and we have independent measurements of the black hole mass $M$, the black hole distance $D$, and the inclination angle of the disk $i$, we can fit the thermal component and infer the black hole spin $a_*$ and the mass accretion rate $\dot{M}$. This is the continuum-fitting method and is normally applied to stellar-mass black holes only: the temperature of the disk indeed scales as $M^{-0.25}$ and the radiation from the inner part of the accretion disk peaks in the soft X-ray band for stellar-mass black holes and in the optical/UV band for the supermassive ones. In the latter case, dust absorption prevents the possibility of accurate measurements. The continuum-fitting method can be extended to test the Kerr hypothesis. If we have a parametric black hole spacetime with one deformation parameter, we can fit the thermal spectrum and infer the black hole spin $a_*$, the mass accretion rate $\dot{M}$, and the deformation parameter.

Here, only to illustrate the {\it potential} capabilities of {\sc nkbb} to constrain the deformation parameter $\alpha_{13}$ from real data, we simulate two observations and we fit the faked data to measure $a_*$, $\dot{M}$, and $\alpha_{13}$. For simplicity, the total spectrum is only the thermal component of the disk and we include the effect of Galactic absorption. In XSPEC language, the total model is {\sc tbabs$\times$nkbb}. The input values of the model parameters are shown in Tab~\ref{tab1} and are similar to the parameters of Cygnus~X-1, with the exception of $a_*$ which is 0.5 in the first simulation and 0.98 in the second simulation. For both simulations, we assume an observation of 200~ks with \textsl{NICER}~\cite{nicer}. When we fit the faked data, all the model parameters are frozen to their input values with the exception of $a_*$, $\dot{M}$, and $\alpha_{13}$, which are free and are determined by the fit.

\begin{figure*}[t]
\begin{center}
\includegraphics[width=8.5cm]{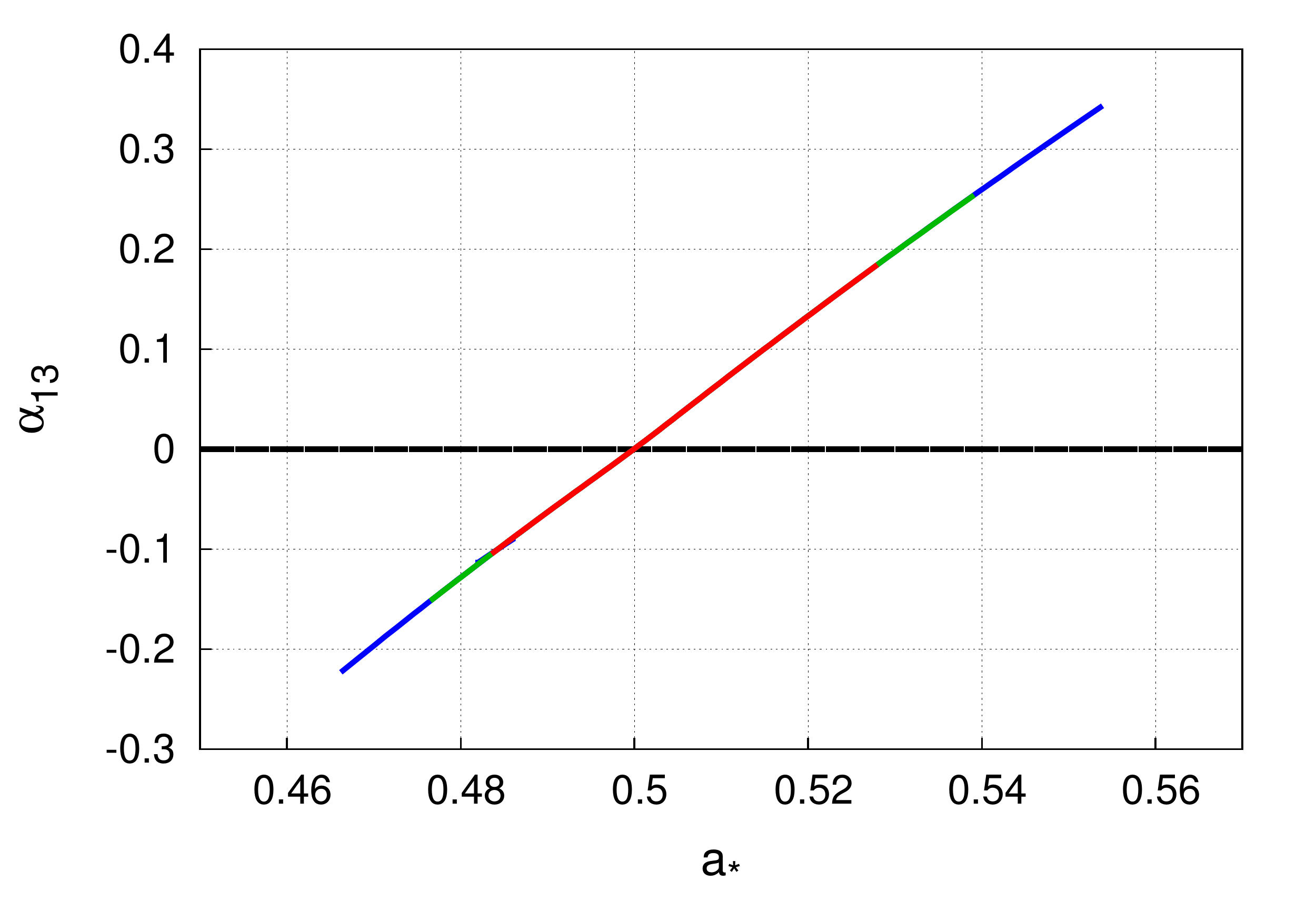}
\includegraphics[width=8.5cm]{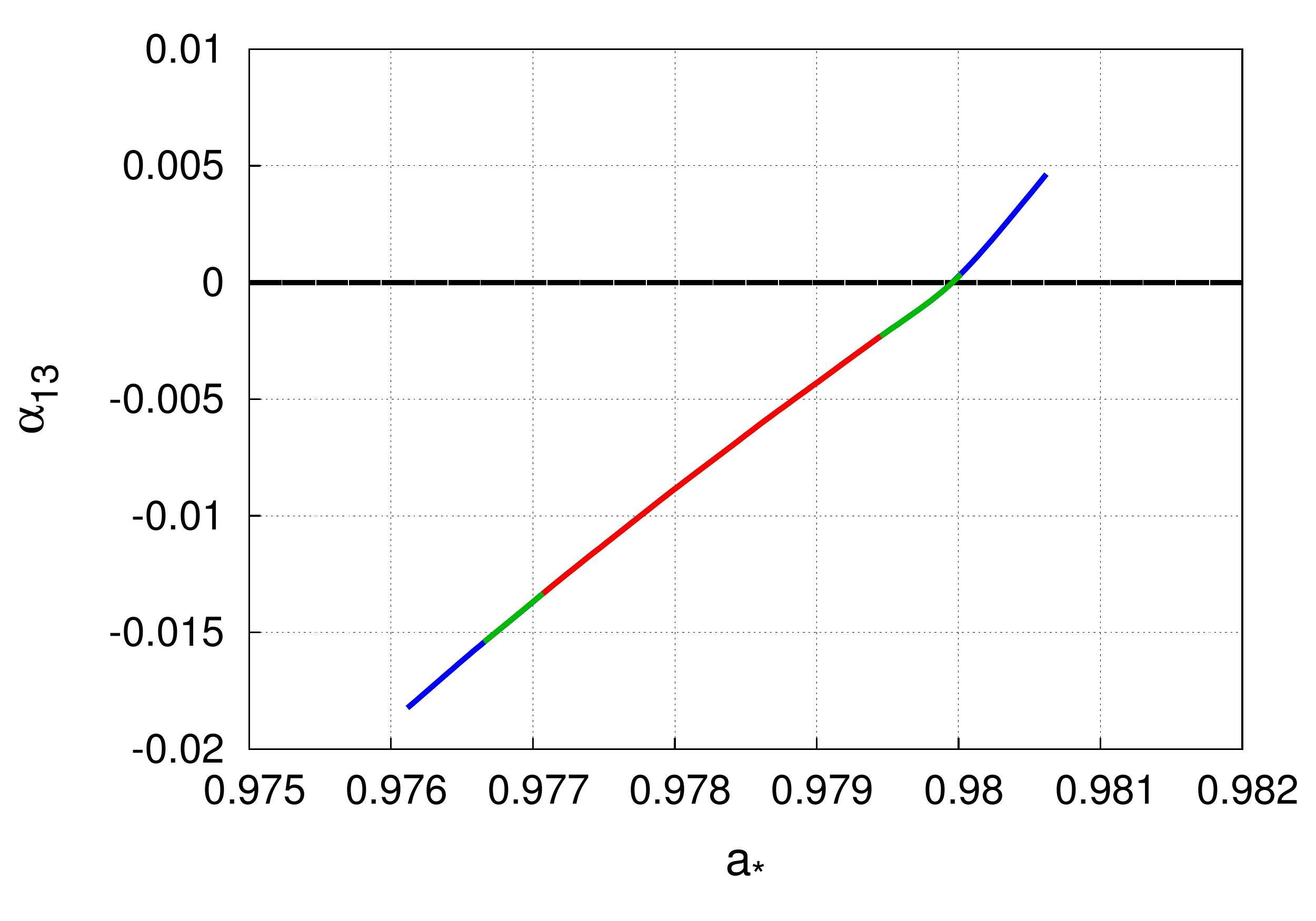}
\end{center}
\vspace{-0.5cm}
\caption{Constraints on the spin parameter $a_*$ and the deformation parameter plane $\alpha_{13}$ from the two simulated observations with \textsl{NICER}. The input parameters are shown in Tab.~\ref{tab1} and the input spin parameter is $a_* = 0.5$ in the left panel and $a_* = 0.98$ in the right panel. The red, green, and blue curves are, respectively, the 68\%, 90\%, and 99\% confidence level curves for two relevant parameters (corresponding to $\Delta\chi^2 = 2.30$, 4.61, and 9.21, respectively). $a_*$ and $\alpha_{13}$ are so strongly correlated that the three confidence level curves appears as lines rather than closed curves. Note also that the difference in the values of the spin parameter in the two simulations leads to an order of magnitude of difference in the constraint on $\alpha_{13}$. \label{f-sim}}
\end{figure*}

We use the $\chi^2$-statistics and the constraints on the spin parameter vs deformation parameter plane that are obtained from the analysis of the two simulated observations are shown in Fig.~\ref{f-sim} (left panel for the simulation with input $a_* = 0.5$, the right panel for that with input $a_* = 0.98$). The red, green, and blue curves are, respectively, the 68\%, 90\%, and 99\% confidence level curves for two relevant parameters (i.e., $\Delta\chi^2 = 2.30$, 4.61, and 9.21, respectively). If one of the two parameters were known, the other could be measured with high precision, because our simplified analysis completely ignores the uncertainties on $M$, $D$, and $i$ and, at the same time, the simulated data have very high statistics. However, in our case we want to determine both parameters by fitting the thermal spectrum of the disk and $a_*$ and $\alpha_{13}$ are strongly correlated. The result is that the three closed confidence level curves look like lines in Fig.~\ref{f-sim}. Note that if the spin is higher we get a stronger constraint on $\alpha_{13}$. Roughly speaking, this is because the inner edge of the accretion disk is closer to the black hole, relativistic effects are stronger, and it is possible to get better estimates of the model parameters. We stress again that Fig.~\ref{f-sim} is only meant to illustrate the potential capability of {\sc nkbb} to test the Kerr hypothesis. In reality, we have to take into account the uncertainties on $M$, $D$, and $i$, which are often known with an accuracy not better than 10\% and eventually they should provide the main contribution to the final uncertainties on the estimates of $a_*$ and $\alpha_{13}$.

\begin{table}
\centering
\vspace{0.3cm}
\begin{tabular}{lc}
\hline
Parameter & \hspace{0.6cm} Input value \hspace{0.6cm} \\
\hline
{\sc tbabs} & \\
$n_{\rm H}$~[$10^{22}$~cm$^{-2}$] & 0.6 \\
\hline
{\sc nkbb} & \\
$M$~[$M_\odot$] & 15 \\
$\dot{M}$~[$10^{18}$~g~s$^{-1}$] & 0.1 \\
$D$~[kpc] & 1.86 \\
$a_*$ & 0.5, 0.98 \\
$\alpha_{13}$ & 0 \\
$i$~[deg] & 30 \\
$f_{\rm col}$ & 1.7 \\
$\Upsilon$ & 1 \\
\hline
\end{tabular}
\vspace{0.3cm}
\caption{Input values of the parameters in the two simulated observations with \textsl{NICER} to illustrate the potentialities to test the Kerr hypothesis with {\sc nkbb}.}
\label{tab1}
\end{table}

%%%%%%%%%%%%%%%%%%%%%%%%%%%%%%%

\section{Concluding remarks \label{s-dis}}

The continuum-fitting method and X-ray reflection spectroscopy are currently the two leading methods to measure black hole spins with electromagnetic techniques under the assumption that the spacetime metric around these objects is described by the Kerr solution of general relativity. Both methods can be extended to probe the spacetime metric around astrophysical black holes and test the Kerr hypothesis. Recently, our group has constructed the XSPEC model {\sc relxill\_nk} to test the Kerr hypothesis using X-ray reflection spectroscopy. In this work, we have presented the XSPEC model {\sc nkbb}, which can be used to test the Kerr hypothesis with the continuum-fitting method. The current version of the model employs the Johannsen metric, but it can be easily extended to any other stationary, axisymmetric, and asymptotically-flat black hole spacetime with an analytic expression of its metric. We have compared our new model with the existing Kerr model {\sc kerrbb} in order to check its accuracy, and we have illustrated the potential capabilities of {\sc nkbb} to constrain the deformation parameter $\alpha_{13}$ with two simulations. In a forthcoming paper, we will apply {\sc nkbb} to analyze the X-ray data of a specific source in order to test the Kerr hypothesis with real observations.

%%%%%%%%%%%%%%%%%%%%%%%%%%%%%%%

{\bf Acknowledgments --}
This work was supported by the Innovation Program of the Shanghai Municipal Education Commission, Grant No.~2019-01-07-00-07-E00035, National Natural Science Foundation of China (NSFC), Grant No.~U1531117, and Fudan University, Grant No.~IDH1512060. A.B.A. also acknowledges the support from the Shanghai Government Scholarship (SGS). S.N. acknowledges support from the Excellence Initiative at Eberhard-Karls Universit\"at T\"ubingen and from the Alexander von Humboldt Foundation.

%%%%%%%%%%%%%%%%%%%%%%%%%%%%%%%

\end{document}